\documentclass{aa}
\usepackage{graphicx}
\usepackage{natbib}
\newcommand{\ROSAT}{{\it ROSAT}}
\newcommand{\Chandra}{{\it Chandra}}
\begin{document}
   \title{The diffuse X-ray background}

   \author{A. M. So\l tan}

   \offprints{A. M. So\l tan}

   \institute{Nicolaus Copernicus Astronomical Center, Bartycka 18,
               00-716 Warsaw, Poland\\
              \email{soltan@camk.edu.pl}% (AMS)\\
             }

   \date{Received ~~~~~~~~~~~~~~~~~; accepted }

  \abstract{
The deepest observations of the X-ray background
approach the surface brightness of the truly diffuse component generated
by Thomson scattering of cosmic X-ray photons. Available estimates
of the electron density and the X-ray luminosity density of AGNs 
as a function of cosmological epoch are used to calculate the
integral scattered X-ray background component. It is shown that
the scattered component constitutes $1.0 - 1.7$\,\% of the total
background, depending on the AGN cosmic evolution. Albeit this is a minute
fragment of the total flux, it becomes a perceptible fraction of the still
unresolved part of the background and should be taken into account
in the future rigorous assessments of the X-ray background structure.
This diffuse component at energies $\lse 1$\,keV sums up with the emission
by WHIM to $3-4$\,\%. Consequently, one should expect that integrated
counts of discrete sources account for just $96-97$\,\%
for soft background and $\sim 99$\,\% at higher energies.
   \keywords{X-rays: diffuse background}
   }
   \authorrunning{A. M. So\l tan}

\maketitle

\section{Introduction \label{intro}}

Most of the X-ray background (XRB) is generated by discrete extragalactic
sources (e.g. %\citealt{lehmann01},
\citealt{campana01}, \citealt{rosati02}, and references therein).
Among those sources, various classes of AGNs constitute a dominating part.
Probably 5 to 10\,\% of the soft XRB is produced by hot gas in clusters of
galaxies. Hydrodynamical simulations of the evolution of the primordial gas
demonstrate that some contribution to the XRB is generated by the baryonic
matter which fills intergalactic space in the form of moderately hot plasma
(\citealt{cen99}, \citealt{dave01}, \citealt{bryan}, \citealt{croft}).
Due to its low density, thermal emission of this gas is extremely weak and the
very existence of the ``warm" baryons still waits for definite observational
confirmation (e.g. \citealt{soltan02}). Thus, apart from this small constituent,
it is generally accepted that the XRB is produced by discrete sources.
This notion has been strengthened directly by deep observations by \ROSAT\
and more recently by \Chandra.

A relationship between the total XRB flux and integrated flux produced
by sources has been investigated in detail from the observational point
by \cite{moretti03}.  These authors compared the source counts with the
measured XRB level in two energy bands: soft ($0.5-2$\,keV) and hard
($2-10$\,keV). They have shown that in the soft band the source counts
integrated down to the \Chandra\ limit of
$2.4\times 10^{-17}$ erg\,s$^{-1}$cm$^{-2}$ produce $94.3^{+7.0}_{-6.7}$\,\%
of the XRB. A smooth extrapolation of counts down to
$\sim\!3\times 10^{-17}$erg\,s$^{-1}$cm$^{-2}$ generates $96$\,\%
of the XRB and is ``consistent with its full value at $1\,\sigma$ level".
In the hard band discrete sources generate at most $93$\,\% (after source counts
extrapolation) and are ``only marginally consistent" with the XRB level.
One should note also that the total XRB flux itself is not well determined;
according to \cite{moretti03} the uncertainties of the XRB in both bands amount
roughly to $5$\,\%.

Although the present data do not allow definite conclusions, it is possible
that the apparent deficiency of the source counts contribution to the hard band
could be removed by steepening of counts below the present threshold of
$2.6\times 10^{-16}$erg\,s$^{-1}$cm$^{-2}$. Such steepening is in fact predicted
by recent evolutionary models of the obscured AGNs (e.g. \citealt{franceschini02}).

Since the deepest observations allow only for a few percent of the XRB in the form
of genuine diffuse radiation, it is worthwhile to examine also those emission
mechanisms which generate a weak signal as compared to the total XRB flux,
but which potentially could contribute significantly to the diffuse component.
The aim of the present paper is to estimate flux of the truly diffuse emission
produced by Thomson scattering of X-rays in intergalactic space. 
The amplitude of the integrated scattered flux depends on
the cosmic history of the XRB and on the density of free electrons as a function
of redshift. These functions are investigated in Sects.~\ref{emissivity} and
\ref{electron}, respectively. In Sect.~\ref{scatter}
the amplitude of the scattered component is calculated and it is discussed what
constraints on the faint end of the source counts are imposed by the present
results.

Most of the AGN related XRB is generated at redshifts smaller
than $\sim 3$ with practically no contribution at redshifts greater than $6$.
Since cumulative Thomson scattering optical depths at redshift $3$ and $6$
do not exceed respectively $\sim 0.02$ and $0.03$ (\citealt{cen03}),
the effects of X-ray absorption remain low and the calculations
below do not include effects of multiple scattering of X-rays.
Integration of the radiative transfer is reduced to calculations of the XRB
intensity as a function of redshift and then integration over redshift of the
scattered component.

\section{AGN luminosity density and the X-ray background \label{emissivity}}

\subsection{Local XRB}

The relationship between the average X-ray background intensity and redshift
is investigated. The XRB flux in the local universe is known with relatively
high accuracy (e.g. \citealt{moretti03}, and reference therein). It is established
that a large fraction of the local XRB is produced by AGNs distributed over
a wide range of redshifts. Due to the strong cosmic evolution of the AGN
population, the XRB varied substantially in the past cosmological epochs
and AGNs constituted a dominating source of the X-ray radiation for most
of the lifetime of the universe. In the present estimates only the AGN
contribution to the XRB is considered, although some other classes of objects
such as young galaxies could contribute to the XRB at high redshifts.

Observed locally at energy $E$ the XRB flux $I_\circ(E)$ is equal to the
integrated flux produced by sources along the line of sight:

\begin{equation}
I_\circ(E) = \int_0^{z_{\rm max}} dz {dV\over dz} {{\cal L}_z(E^\prime)\cdot
             (1+z)\over 4\pi D_L^2},                           \label{local_bkg}
\end{equation}
where $V$ is the co-moving volume, $D_L$ is the luminosity distance,
and ${\cal L}_z(E^\prime)$ is the luminosity density generated by AGNs at
redshift $z$ and energy $E^\prime = E\cdot(1+z)$.  The maximum redshift,
$z_{\rm max}$, is the redshift at which the first AGNs began
to shine. Systematic variations of the luminosity density with redshift are
described by the cosmic evolution ${\cal E}(z)$:

\begin{equation}
{\cal L}_z = {\cal L}_o \cdot {\cal E}(z).                     \label{lum_dens}
\end{equation}

Various models for ${\cal E}(z)$ have been proposed in the literature.
The models differ between each other in details, but the general shape of the
evolution is common. It is accepted that at low redshifts the luminosity density
generated by AGNs rises sharply, stabilizes at moderate redshifts and probably
decreases at high redshifts. One should note that in the present investigation the
integrated luminosity density as a function of redshift is the only ``interesting"
quantity. Thus, neither the exact shape of the X-ray luminosity function nor
the evolution type ({\it luminosity} vs. {\it density} evolution) affect
the calculations. It was found that the \cite{miyaji00} estimates of the AGN
luminosity function and cosmological evolution can be used to determine
the luminosity density in a straightforward and effective way.
Their {\it Luminosity Dependent Density Evolution} model (designated as LDDE1)
is applied. It is based on the combined shallow and deep \ROSAT\ surveys
and adequately reproduces both the source counts and the AGN redshift distribution. 

\begin{figure}
\centering
\includegraphics[width=0.9\linewidth]{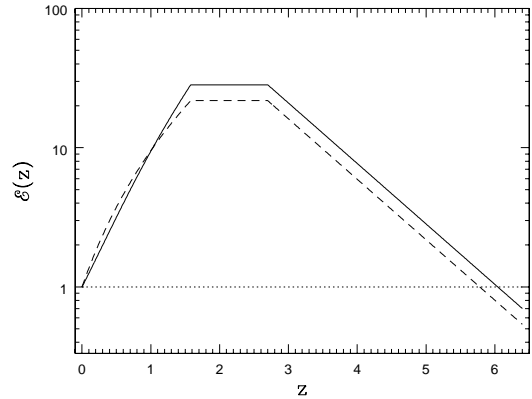}
\caption{Rate of the cosmological evolution of the luminosity density
according to the LDDE1 model with a decline given in Eq.~(3)
by \cite{miyaji00} (solid) and density evolution by \cite{boyle94} (dashed).}               \label{evolution}
\end{figure}

The present calculations cover a wide range of redshifts. The observed
energy band [$E_1,E_2$] corresponds to the band [$E_1(1+z),E_2(1+z)$] at the
source. Since the observed flux results from the integrated emission
along the line of sight, the relevant luminosity function and evolution are
defined in the (redshifted) source frame (see discussion in \cite{miyaji00}).
In the paper the soft X-ray band $0.5-2$\,keV is used as reference and all
the data extracted from the literature are corrected to this standard band.

The AGN sample
used in the \cite{miyaji00} analysis contained a limited number of high
redshift QSOs ($17$ with $z> 2.2$). In effect, constraints on the evolution
rate at high redshifts are not restrictive. \cite{miyaji00} note, however,
that the X-ray data are consistent with the decline of the space density
of optically selected QSO for redshifts greater than $\approx 2.7$.
To account for this trend, the evolution function, ${\cal E}(z)$, is
calculated from the LDDE1 model with an additional ``damping factor":
\begin{equation}
{\cal E}_z\; \sim\; \exp(2.7-z) \;\; {\rm for}\;\; z>2.7.
                                                         \label{damping}
\end{equation}
The cosmological model $(\Omega_m,\Omega_\Lambda) = (0.3, 0.7)$ has been used
in all the calculations. In Fig.~\ref{evolution} the evolution rate defined
as a ratio of the co-moving luminosity density at redshift $z$ to
its local value is shown for the LDDE1 model. The density evolution derived by
\cite{boyle94} is plotted for comparison. Differences between both evolution
laws of $\sim 25$\,\% which arise at $z\gse 1.5$ indicate the level of
uncertainties involved in the evolution estimates (see below).

A relationship between volume, distance and redshift for the cosmological
model used in the paper is taken from \cite{hogg99}:

\begin{equation}
{dV\over dz}\,{1\over D_L^2} = \omega {c\over H_\circ}\,
                       {1\over F_\circ(z)\,(1+z)^2}\;,       \label{dvdz}
\end{equation}
where
\begin{equation}
F_\circ(z) = \sqrt{\Omega_m(1+z)^3 + \Omega_\Lambda}          \label{e0}
\end{equation}
with $\Omega_m = 0.3$ and $\Omega_\Lambda = 0.7$. Eqs.~(4) and (5)
inserted into Eq.~(1) give for unit solid angle $\omega = 1$\,sr:
\begin{equation}
I_\circ(E) = {1\over 4\pi}\,{c\over H_\circ}\,{\cal L}_\circ(E)
             \int_0^{z_{\rm max}} dz\,{{\cal E}(z)\over F_\circ(z)\,(1+z)^2}\;.
                                                              \label{local_bkg_1}
\end{equation}

The local background flux $I_\circ(E)$ is only weakly dependent on the maximum
redshift in the integral and in all the calculations $z_{\rm max} = 6$ was put.
The local luminosity density and evolution have been computed from
the \cite{miyaji00} formulae and inserted into Eq.~(6). The results
are dependent on extrapolation of the X-ray luminosity function (XLF)
at the low luminosity end. In the present calculations the XLF is
integrated over the luminosity range actually covered by the \cite{miyaji00}
sample, i.e. between $L_{\rm min} = 10^{41.5} h_{50}^{-2}$ erg\,s$^{-1}$
and $L_{\rm max} = 10^{47} h_{50}^{-2}$erg\,s$^{-1}$,
where $h_{50}=H_\circ / 50$\,km\,s$^{-1}$Mpc$^{-1}$. In the energy
band of $0.5-2$\,keV the AGNs according to the LDDE1 with the decline
described by Eq.~(3) model produce
a background flux of $1.97\times 10^{-8}$\,erg\,s$^{-1}$\,cm$^{-2}$\,sr$^{-1}$.
Using the \cite{moretti03} estimate of the total soft XRB at $2.47\times 10^{-8}$
erg\,s$^{-1}$\,cm$^{-2}$\,sr$^{-1}$, it gives the AGN contribution at
$\sim 80$\,\%. A pure LDDE1 model (without decline of luminosity density at
high redshifts) generates a flux of
$2.16\times 10^{-8}$\,erg\,s$^{-1}$\,cm$^{-2}$\,sr$^{-1}$ ($= 88$\,\% of the
total XRB).

\begin{figure}
\centering
\includegraphics[width=0.9\linewidth]{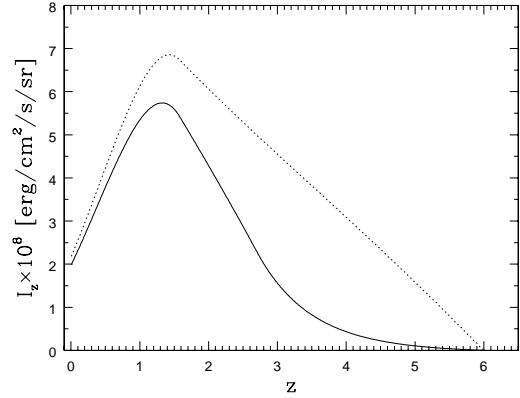}
\caption{XRB flux in the band $0.5(1+z)$keV - $2.0(1+z)$keV per co-moving
cm$^2$ measured at redshift $z$. The solid curve represents data for the LDDE1
model with decreasing luminosity density according to Eq.~(3), the
dotted curve - the LDDE1 model without ``damping". In both cases the luminosity
density is set to zero for $z>6$.}               \label{bkg_z}
\end{figure}

\subsection{XRB in the past}

A hypothetical observer at redshift $z$ would detect the integrated XRB
appropriate for his%\footnote{or her}
~epoch:
\begin{equation}
I_z^\prime(E^\prime) =
         \int_0^{z_{\rm max}^\prime} dz^\prime\,{dV^\prime\over dz^\prime}
         \,{{\cal L}_{z_1}^\prime[E^\prime\,(1+z^\prime)]\cdot (1+z^\prime)\over
         4\pi {D_L^\prime}^2}\;,                                   \label{z_bkg}
\end{equation}
where primes ($^\prime$) indicate that all quantities refer to and are measured
by the observer at redshift $z$. Photons emitted at redshift $z^\prime$ were
sent at an epoch, which for us corresponds to the redshift $z_1$:
\begin{equation}
1+z_1 = (1+z)\cdot (1+z^\prime)\;.                       \label{z1zzprime}
\end{equation}
The cosmological parameters in Eq.~(7) satisfy relationships analogous
to Eqs.~(4) and (5):
\begin{equation}
{dV^\prime\over dz^\prime}\,{1\over {D_L^2}^\prime} = \omega {c\over H_z}\,
              {1\over F_z(z^\prime)\,(1+z^\prime)^2}\;,       \label{dvdzprime}
\end{equation}
and
\begin{equation}
F_z(z^\prime) =
      \sqrt{\Omega_m^\prime(1+z^\prime)^3 + \Omega_\Lambda^\prime}\;.   \label{ez}
\end{equation}
Standard relationships (e.g. \citealt{hogg99}) give:
\begin{equation}
H_z = H_\circ\cdot F_\circ(z)\;,
\end{equation}
\begin{equation}
F_\circ(z_1) = F_\circ(z)\cdot F_z(z^\prime)\;,
\end{equation}
\begin{equation}
{\cal L}_{z_1}^\prime = (1+z)^3 {\cal L}_{z_1}\;\,
\end{equation}
\begin{equation}
I_z(E^\prime) = {1\over (1+z)^2}\,I_z^\prime(E^\prime)\;, \label{izizprime}
\end{equation}
where $I_z^\prime(E^\prime)$ is the XRB flux per proper unit surface measured
by observer at redshift $z$ and $I_z(E^\prime)$ represents flux
per co-moving unit surface. Using Eqs.~(8) -- (14) we finally get:
\begin{eqnarray}
\lefteqn {
J_z(E_1^\prime,E_2^\prime) = } \nonumber \\
 & & {c\over H_\circ} (1+z) {\cal L}_\circ(E_1,E_2)
                \int_0^{z_{\rm max}^\prime} dz^\prime {{\cal E}_\circ(z_1)
                \over F_\circ(z_1)(1+z^\prime)^2}\;,
                                                       \label{jz}
\end{eqnarray}
where $J_z(E_1^\prime,E_2^\prime)$ is the XRB flux per co-moving unit surface
measured at redshift $z$ within the energy band $E_1^\prime-E_2^\prime$
multiplied by a solid angle of $4\pi$.

\section{Electron density in intergalactic space \label{electron}}

Extensive hydrodynamical simulations of the matter distribution
(e.g.\citealt{valageas02} and references therein) show that a dominating
fraction of all the baryonic matter in the universe resides outside galaxies
and cluster of galaxies. In the local universe baryons accumulated
in these objects constitute $\sim 30$\,\% of the total baryonic mass, while
the remaining $\sim 70$\,\% still stays in the {\it diffuse} and
{\it warm-hot} phases filling $99$\,\% of volume space. The fraction of the
diffuse components rises with redshift and reaches roughly $99$\,\% at $z=3$
(\citealt{cen99}, \citealt{dave01}). Intergalactic matter is virtually
fully ionized up to $z\approx6$ (e.g. \citealt{cen03}).

Estimates of the total baryonic mass based on {\it WMAP} observations
give $\Omega_b = 0.044$ for $H_\circ = 71$\,km\,s$^{-1}$\,Mpc$^{-1}$
(\citealt{spergel03}). Using results on the fraction of baryons in the
intergalactic space by \cite{cen99}, the {\it WMAP} data give the local
average electron density at $n_e(0) = 2.2\times 10^{-7}$\,cm$^{-3}$.
Due to systematic increase of the diffuse component with redshift,
the electron co-moving density rises to $n_e(z) = 2.2\times 10^{-6}$\,cm$^{-3}$
at $z=1$ and to $n_e(z) = 2.0\times 10^{-5}$\,cm$^{-3}$ at $z=3$. At
still higher redshifts, practically all baryons reside in the diffuse
component and $n_e(z) \sim (1+z)^3$.

\section{The scattered component \label{scatter}}

The luminosity density generated by the Thomson scattering of the XRB
is given by:
\begin{equation}
{\cal L}_z^T = n_e(z)\: \sigma_T\: J_z\;,            \label{scat_lum}
\end{equation}
where $\sigma_T = 6.65\times 10^{-25}$\,cm$^2$ is the Thomson
cross-section. The XRB flux due to Thomson scattering is calculated
using an equation analogous to Eq.~(6):
\begin{equation}
I_\circ^T(E_1,E_2)= {1\over 4\pi}\,{c\over H_\circ} \int_0^{z_{\rm max}}
           dz\,{{\cal L}_z^T(E_1^\prime,E_2^\prime)\over
           F_\circ(z)\,(1+z)^2}\;.
\end{equation}

Due to the strong dependence of the electron density on redshift, the scattered
component is generated primarily at high redshifts. The standard LDDE1 model 
with constant luminosity density for $1.6 < z < 6$ represented by the dotted
curve in Fig.~\ref{bkg_z} generates scattered XRB flux of 
$4.14\times 10^{-10}$\,erg\,s$^{-1}$\,cm$^{-2}$\,sr$^{-1}$, which accounts for
$1.7$\,\% of the total XRB. If the AGN luminosity density declines exponentially
for redshifts $z>2.7$ according to Eq.~(3) (solid curve
in Fig.~\ref{bkg_z}), the scattered XRB flux is reduced to
$2.42\times 10^{-10}$\,erg\,s$^{-1}$\,cm$^{-2}$\,sr$^{-1}$ or
$0.98$\,\% of the total XRB.

Low optical depth for Thomson scattering for $z<6$ implies relatively weak
signal of the scattered XRB component. However, in the deepest exposures of
the present-day instruments just a few percent of the soft XRB remain
unresolved into discrete sources. Thus, the scattered component contributes
measurably to the unresolved part. Moreover, a separate diffuse contribution
by the WHIM is expected in the soft XRB. The amplitude of this component has
not been determined yet with reasonable accuracy (\citealt{soltan02}).
The WHIM contribution to the XRB is expected to drop sharply with redshift
(\citealt{cen99}) and it is likely that thermal emission in the $0.5-2.0$\,keV
band does not exceed $2-3$\,\% of the total XRB. \cite{moretti03} point out
that a smooth extrapolation of the counts down to a flux level of $\sim 3\times
10^{-18}$\,erg\,s$^{-1}$\,cm$^{-2}$\,sr$^{-1}$, i.e. a factor of 10
below the present limits could account for $96$\,\% of the total XRB.
Although this estimate is subject to relatively high uncertainties
($\sim 7$\,\% at the $1\,\sigma$ level), the existence of the diffuse
component further strengthens the conclusion that
with a moderate extrapolation of source counts, all constituents
of the XRB are accounted for and there is no room for new classes of
X-ray sources. In particular, in the soft band the source counts cannot
exhibit steepening which would increase the integrated XRB more than
a few percent.

In the harder band of $2-10$\,keV source counts extrapolated down to fluxes
an order of magnitude below the present threshold generate $\sim 93$\,\% 
of the total XRB (\citealt{moretti03}). The relative contribution of the
Thomson scattered component in this band is similar to the contribution
in soft X-rays, while the expected WHIM emission above $2$\,keV is negligible.
Thus, taking into account the diffuse component does not change the
\cite{moretti03} conclusion that the source counts are likely to
steepen below $\sim 10^{-15}$\,erg\,s$^{-1}$\,cm$^{-2}$
due to a population of highly obscured, hard sources.

Taking into account the uncertainties of the absolute flux of the total XRB
and uncertainties of the discrete source contribution, which both are at
the level of several percent, calculations in the present paper do not address
the question whether the source counts match exactly the observed XRB. One
should stress that the objective of the present investigation was to estimate
the amount of the diffuses XRB component generated by Thomson scattering.
The amplitude of this component is calculated using the AGN X-ray luminosity
function and evolution. The uncertainty of the diffuse component flux results
mainly from uncertainties related to the evolution of the AGN population. The
present calculations show that the Thomson scattered flux amounts to or slightly
exceeds $1$\,\% of the total XRB, Thus, as long as uncertainties of the relevant
measurements are subject to larger errors, this component does not
affect decisively the XRB budget. Nevertheless, in the soft band the Thomson
component narrows noticeably the distance between the resolved and total XRB
and reduces the range of the allowed $\log S - \log N$ relatioships.
One might expect that the relative importance of the Thomson component will
grow with a further increase of the resolved fraction of the background.

Direct detection of the scattered component poses a serious observational
problem. This is because the spectral shape of the scattered flux strictly
corresponds to the average AGN spectrum and the diffuse component mimics
a population of unresolved AGN-like sources. The amplitude of the diffuse
flux could be determined only indirectly by subtraction of the discrete
source contribution from the total XRB. In practical terms, to establish
precisely the flux produced by discrete sources one needs to determine
the $\log N - \log S$ relationship well below the present limits.

\vspace{2mm}
ACKNOWLEDGMENTS.
This work has been partially supported by the Polish KBN grants 5~P03D~022~20
and PBZ-KBN-054/P03/2001.


\begin{thebibliography}{}

\bibitem[Boyle et al.\ (1994)]{boyle94}
   Boyle, B. J., Shanks, T., Georgantopoulos, I, Stewart, G. C.,
   Griffiths, R. E., 1994, MNRAS, 271, 639

\bibitem[Bryan \& Voit (2001)]{bryan}
   Bryan, G. L. \& Voit, G. M., 2001, ApJ, 556, 590

\bibitem[Campana et al.\ (2001)]{campana01}
   Campana, S., Moretti, A., Lazzati, D. \& Tagliaferri, G., 2001,
   ApJ, 560, L19

\bibitem[Cen (2003)]{cen03}
   Cen, R., 2003, ApJ, 591, L5

\bibitem[Cen \& Ostriker(1999)]{cen99}
   Cen, R. \& Ostriker, J. P., 1999, ApJ, 514, 1

\bibitem[Croft et al.\  (2001)]{croft}
   Croft, R. A. C., Di Matteo, T., Dav\'e, R., et al., 2001, ApJ, 557,67

\bibitem[Dav\'e et al.\  (2000)]{dave01}
   Dav\'e, R., Cen, R., Ostriker, J. P., et al., 2000, ApJ, 552, 473

\bibitem[Franceschini et al.\ (2002)]{franceschini02}
    Franceschini, A., Braito, V., Fadda, D., 2002, MNRAS, 335, L51

\bibitem[Hogg (1999)]{hogg99}
    Hogg, D. W., 1999, {\it astro-ph/9905116}

%\bibitem[Lehmann et al.\  (2001)]{lehmann01}
%   Lehmann, I., Hasinger, G., Schmidt, M., et al., 2001, A\&A, 371, 833

\bibitem[Miyaji et al.\ (2000)]{miyaji00}
   Miyaji, T., Hasinger, G., Schmidt, M., 2000, A\&A, 353, 25

\bibitem[Moretti et al.\ (2003)]{moretti03}
   Moretti, A., Campana, S., Lazzati, D., Tagliaferri, G., 2003, ApJ, 588, 696

\bibitem[Rosati et al.\ (2002)]{rosati02}
   Rosati, P., Tozzi, P., Giacconi, R., et al., 2002, ApJ, 566, 667

\bibitem[So\l tan et al.\ (2002)]{soltan02}
   So\l tan A. M., Freyberg, M., Hasinger G., et al., 2002, A\&A, 395, 475

\bibitem[Spergel et al.\ (2003)]{spergel03}
   Spergel, D. N., Verde, L., Peiris, H. V., 2003, {\it astro-ph/0302209}, ApJ accepted

\bibitem[Valageas et al.\ (2002)]{valageas02}
   Valageas, P., Schaeffer, R., Silk, J., 2002, A\&A, 388, 741

\end{thebibliography}
\end{document}